\documentclass[aps,prl,reprint,amsmath,amssymb,superscriptaddress]{revtex4-2}

\usepackage[utf8]{inputenc}
\usepackage[T1]{fontenc}

\usepackage{dsfont}
\usepackage{graphicx} 
\graphicspath{{figures/}}

\usepackage{amsmath}
\usepackage{amsfonts}
\usepackage{amssymb}
\usepackage{physics}
\usepackage[table]{xcolor}  

\usepackage{booktabs}    
\usepackage{pifont}

\usepackage[colorlinks=true,linkcolor=blue,urlcolor=blue,citecolor=blue,pdfusetitle]{hyperref}

\newcommand{\cmark}{\textcolor{green!50!black}{\checkmark}}
\newcommand{\xmark}{\textcolor{red!50!black}{\ding{55}}}

\begin{document}

\title{Spin-orbit-enabled realization of arbitrary two-qubit gates on moving spins}

\author{D.~Fernández-Fernández}
\email{david.fernandez@csic.es}
\affiliation{Instituto de Ciencia de Materiales de Madrid, CSIC, Cantoblanco, E-28049 Madrid, Spain}
\author{Y.~Matsumoto}
\affiliation{QuTech and Kavli Institute of Nanoscience, Delft University of Technology, Delft, The Netherlands}
\author{L.M.K.~Vandersypen}
\affiliation{QuTech and Kavli Institute of Nanoscience, Delft University of Technology, Delft, The Netherlands}
\author{G.~Platero}
\affiliation{Instituto de Ciencia de Materiales de Madrid, CSIC, Cantoblanco, E-28049 Madrid, Spain}
\author{S.~Bosco}
\affiliation{QuTech and Kavli Institute of Nanoscience, Delft University of Technology, Delft, The Netherlands}

\begin{abstract}
Shuttling spin qubits in systems with large spin-orbit interaction (SOI) can cause errors during motion.
However, in this work, we demonstrate that SOI can be harnessed to implement an arbitrary high-fidelity two-qubit (2Q) gate.
We consider two spin qubits defined in a semiconductor double quantum dot that are smoothly moved toward each other by gate voltages.
We show that an arbitrary high-fidelity 2Q gate can be realized by controlling the shuttling speed and waiting times, and leveraging strong intrinsic or extrinsic SOI.
Crucially, performing 2Q operations during qubit transport enables a one-step realization of a wide range of 2Q gates, which often involve several steps when implemented using static dots.
Our findings establish a practical route toward direct implementation of any 2Q gate via spin shuttling, significantly reducing control overhead in scalable quantum computing architectures.
\end{abstract}

\date{\today}

\maketitle

{\bf \emph{Introduction.--}}
Quantum computing promises advantages over classical computation~\cite{Zhong2020, Wu2021}, but practical algorithms demand large, highly-connected quantum chips~\cite{Bravyi2024, Goto2024}.
Semiconductor quantum dots (QDs) are leading candidates for scalable architectures~\cite{Vandersypen2019,Gyoergy2022,Burkard2023}, offering long coherence times~\cite{Hansen2022,Foster2025,Fuentes2025}, high-fidelity gates~\cite{Xue2022,Noiri2022}, and CMOS compatibility~\cite{Maurand2016,Zwerver2022,Geyer2024,Huckemann2025}.
A key challenge remains maintaining high fidelities in dense arrays.
Modular architectures address this by connecting small, efficient units through quantum links~\cite{Vandersypen2017,Kuenne2024,Ginzel_2024}.
Recent experiments have shown coherent spin-qubit shuttling across QD arrays~\cite{Zwerver2023,RiggelenDoelman2024,Struck2024,DeSmet2025,Foster2025}, enabling quantum buses for quantum information transfer.
A key mechanism for spin-qubit control is the spin-orbit interaction (SOI), allowing all-electrical manipulation via electric dipole spin resonance~\cite{Nowack2007,FernandezFernandez_2023,John2024}.
However, SOI combined with noise can introduce decoherence and path-dependent dynamics, degrading shuttling fidelity.
Recent studies show SOI can enhance shuttling by filtering low-frequency noise~\cite{Bosco2024} and enabling qubit control en route~\cite{FernandezFernandez2024,Wang2024}.

\begin{figure}[pt!]
	\centering
	\includegraphics[width=\linewidth]{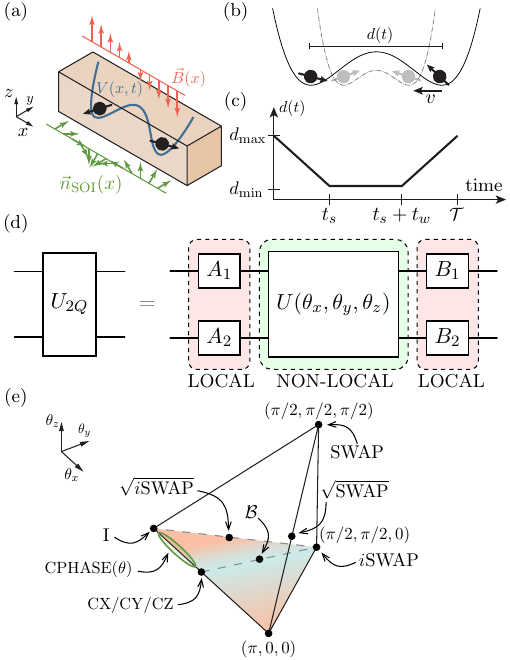}
	\caption{(a) Schematic of two spin qubits confined in a double QD potential $V(x, t)$.
	(b) Conveyor-mode potential $V(x, t)$, where the interdot distance is $d(t)$ and the shuttling speed is $v$.
	(c) The shuttling protocol starts with two spin qubits separated by $d_\mathrm{max}$, and last for a total time $\mathcal{T} = 2 t_s + t_w$.
	(d) Khaneja-Glaser decomposition of a general two-qubit gate into one-qubit local gates (pink) $A_i$ and $B_i$, and a single two-qubit nonlocal gate (green) $U(\theta_x, \theta_y, \theta_z)$.
	(e) Weyl chamber in the space spanned by $\theta_i$.}
	\label{fig:schematic}
\end{figure}

Efficient two-qubit (2Q) gates are essential for scalability.
SOI-enabled 2Q gates have been explored in static arrays~\cite{Geyer2024,Spethmann2024,Qi2024, Bosco2024a, RimbachRuss2024,Nguyen2025}, but implementing them during shuttling, rather than through complex sequences of static operations, can reduce circuit depth and cumulative error.
Direct access to 2Q gates in motion suppresses computation overhead.
Although conveyor-mode~\cite{Seidler2022,Langrock2023,Xue2024,Struck2024,DeSmet2025,Nagai2025} spin shuttling has recently been used to implement high-fidelity 2Q gates~\cite{Matsumoto2025}, the role of SOI in enabling 2Q gates during motion remains unresolved.

In this Letter, we show that large SOI, intrinsic or engineered, enables an arbitrary high-fidelity 2Q gate during spin shuttling.
By tuning only the shuttling speed and waiting time of the conveyor mode, we implement a broad class of entangling gates, not only the typical $\mathrm{CPHASE}$ and $\mathrm{SWAP}$ families, but also fermionic simulation~\cite{Kyriienko2021,Nguyen2024} and Berkeley gates~\cite{Zhang2004}, which are challenging to reach in a single step in current spin-based devices.
Our general framework includes position-dependent magnetic fields, SOI, and anisotropic $g$-factors, and applies to electrons and holes in, e.g., Si, Ge, GaAs and InAs, in planar and nanowire geometries.
Our results provide a realistic and scalable pathway toward long-range 2Q gates with minimal overhead for quantum simulation, distributed computing, and shuttling-based quantum error correction \cite{Siegel2025}.

{\bf \emph{Model.--}}
We consider two spin qubits confined in a double QD potential $V(x, t)$ moving in a one-dimensional path along the $x$-axis, see Fig.~\ref{fig:schematic}~(a).
The Hamiltonian~\cite{Bosco2024}
\begin{equation}
	\begin{split}
		H = \sum_{i=1}^{2}& \left[\frac{p_i^2}{2m^*} + \left\{\vec{v}_\mathrm{SOI}(x_i), p_i\right\}\cdot \vec{\sigma}^{(i)} + \frac{\mu_B \overline{g}}{2} \vec{B}(x_i)\cdot \vec{\sigma}^{(i)}\right. \\
		&\;\left. \vphantom{\frac{p_i^2}{2m^*}} + V(x_i, t) \right] + \frac{1}{4\pi \epsilon}\frac{e^2}{\abs{x_1 - x_2}},
	\label{eq:total_hamiltonian}
	\end{split}
\end{equation}
includes kinetic energy, SOI, magnetic field, confinement and Coulomb repulsion, respectively.
We introduce the momentum acting on the $i$-th particle as $p_i = -i\hbar\partial_{x_i}$, the effective mass $m^*$, the SOI $\vec{v}_\mathrm{SOI}(x_i)=v_\mathrm{SOI}(x_i)\hat{n}_\mathrm{SOI}(x_i)$, with velocity $v_\mathrm{SOI} = \hbar / m^* l_\mathrm{SOI}$ and $l_\mathrm{SOI}$ being the SOI length, the Pauli matrices $\vec{\sigma}^{(i)}$, and the normalized anticommutator $\{\cdot, \cdot\}$.
The Bohr magneton is $\mu_B$, $\overline{g}$ is the effective $g$-tensor, $\vec{B}(x_i)$ is the magnetic field, and $\epsilon$ is the dielectric constant of the medium.
For the sake of concreteness, we focus on Si devices with $m^* = 0.1 m_e$, $\epsilon = 11\epsilon_0$, and isotropic tensor $\overline{g} = g^*\delta_{ij} = 2$, noting that materials with spacially-varying $\overline{g}$, e.g., holes in Ge, can be incorporated by rescaling $\vec{B}(x)$.

The conveyor-mode potential is modeled as
\begin{equation}
	V(x, t) = \frac{1}{2}m^*\omega_0^2\frac{\left[x^2-4d(t)^2\right]^2}{16d(t)^2},
\end{equation}
where $\omega_0$ is the confinement frequency and $d(t)$ the time-dependent interdot distance.
This form ensures constant confinement throughout shuttling, see Fig.~\ref{fig:schematic}~(b).
The confinement length scale reads $l_0 = \sqrt{\hbar / m^* \omega_0}$.

Computing the time-evolution operator of Eq.~\eqref{eq:total_hamiltonian} is challenging, even with split-operator methods~\cite{FillionGourdeau2012,Glowinski2017,Choi2019,Roulet2021}.
We thus derive a low-energy Hamiltonian acting on the computational basis $\{\ket{\uparrow \uparrow}, \ket{\uparrow \downarrow}, \ket{\downarrow \uparrow}, \ket{\downarrow \downarrow}\}$~\cite{Hetenyi2020,Geyer2024,SaezMollejo2025}:
\begin{equation}
	\tilde{H} = \frac{1}{2}\left(\vec{\Delta}_L \cdot \vec{\sigma}_L + \vec{\Delta}_R \cdot \vec{\sigma}_R\right) + \frac{1}{4} \vec{\sigma}_L \cdot \overline{J} \vec{\sigma}_R,
	\label{eq:effective_hamiltonian}
\end{equation}
with $\vec{\Delta}_i = \mu_B \overline{g}_i \vec{B}(x_i)$ capturing local Zeeman fields.
The kinetic energy, SOI, and Coulomb interaction contribute to the exchange matrix $\overline{J}$, which can be written as $\overline{J}=J_0R(\theta, \phi, \alpha)$, where $R$ is a rotation matrix around a vector parametrized by polar angle $\theta$ and azimuthal angle $\phi$, and rotation angle $\alpha$.
These parameters vary with the interdot distance, and thus evolve during shuttling, unlike static 2Q gates.
This effective model is accurate for linear-in-momentum SOI~\cite{Hetenyi2022}.
Further numerical details on $\overline{J}(t)$ are given in the Supplemental Material (SM)~\cite{SupMat}.

The shuttling 2Q gate protocol, as in~\cite{Matsumoto2025}, consists of a linear ramp reducing $d(t)$ from $d_\mathrm{max}$ to $d_\mathrm{min}$ at constant speed $v$, lasting a time $t_s = (d_\mathrm{max} - d_\mathrm{min})/2v$.
After a waiting time $t_w$, the qubits are separated back at the same speed.
The full protocol lasts $\mathcal{T} = 2t_s + t_w$, see Fig.~\ref{fig:schematic}~(c).

{\bf \emph{Two-qubit gates.--}}
To benchmark the 2Q shuttling gate, we note that any 2Q gate ($U_\mathrm{2Q}$) decomposes into single-qubit rotations and a nonlocal 2Q gate, see Fig.~\ref{fig:schematic}~(d).
Explicitly, $U_\mathrm{2Q} = A_1 \otimes B_1 \cdot U(\theta_x, \theta_y, \theta_z) \cdot A_2 \otimes B_2$, where $A_i$ and $B_i$ are single-qubit gates on qubit $i$, and $\cdot \otimes \cdot$ denotes the tensor product.
This is known as the Cartan, or Khaneja-Glaser decomposition~\cite{Khaneja2001,Bremner2002,Tucci2005}.
Because high-fidelity single-qubit gates can be implemented before and after shuttling, we focus on the nonlocal 2Q gate implemented during motion.
Two gates $U_{2Q}$ and $U'_{2Q}$ are locally equivalent, denoted $U_{2Q} \sim U'_{2Q}$, if they differ only by single-qubit operators.

The nonlocal gate reads
\begin{equation}
	\begin{split}
	U(\theta_x, \theta_y, \theta_z) \equiv \tilde{U}_\mathrm{2Q} = &\exp\left[i(\theta_x \sigma_x\otimes \sigma_x + \theta_y \sigma_y\otimes \sigma_y \right.\\
	& \qquad \left. + \theta_z \sigma_z\otimes \sigma_z)\right] ,
	\end{split}
	\label{eq:two_qubit_gate}
\end{equation}
and the tilde denotes the nonlocal component of the 2Q gate.
This decomposition is general, and includes mixed interactions, e.g., $(\sigma_x\otimes \sigma_y + \sigma_y\otimes \sigma_x)$, by single-qubit operations.
Extracting $\theta_i$ for a general 2Q gate is challenging, involving a 15-parameter optimization.
A more effective method uses Makhlin invariants~\cite{Makhlin2002,CalderonVargas2015,Cayao2020},
\begin{equation}
		G_1 = \frac{\Tr^2 m}{16 \det U_{2Q}} \in \mathds{C}, \quad G_2 = \frac{\Tr^2 m - \Tr m^2}{4 \det U_{2Q}} \in \mathds{R},
\label{eq:makhlin_invariants}
\end{equation}
where $m \equiv U_B^T U_B$ and $U_B \equiv Q^\dagger U_{2Q} Q$, with $Q$ the Bell-basis transformation.
If $G_1(U_{2Q})=G_1(U'_{2Q})$ and $G_2(U_{2Q}) = G_2(U'_{2Q})$, then $U_{2Q} \sim U'_{2Q}$.

To recover $\theta_i$, we numerically solve
\begin{subequations}
	\begin{align}
	\begin{split}
		G_1(U_{2Q}) &= \frac{1}{4}e^{-4i\theta_z}\left\{e^{4i\theta_z}\cos[2(\theta_x - \theta_y)] \right. \\
		& \hspace{5em} \left. \vphantom{e^{-4i\theta_z}} + \cos[2(\theta_x + \theta_y)]\right\}^2,
	\end{split}\\
	\begin{split}
		G_2(U_{2Q}) &= \cos(4\theta_x) + \cos(4\theta_y) + \cos(4\theta_z).
	\end{split}
	\end{align}
\end{subequations}
The solutions are not unique for $\theta_i \in [0, \pi]$, so we restrict to the first Weyl chamber~\cite{Zhang2003,Satoh2022}.
This chamber is the tetrahedron defined by vertices $(\theta_x, \theta_y, \theta_z) = (0, 0, 0)$, $(\pi, 0, 0)$, $(\pi/2, \pi/2, 0)$, and $(\pi/2, \pi/2, \pi/2)$, see Fig.~\ref{fig:schematic}~(e).
Restricting to the Weyl chamber ensures that two gates with different $\theta_i$ are not locally equivalent.
This unique representation holds for the entire Weyl chamber, except on the base.
Gates with $(\theta_x, \theta_y, 0)$ and $(\pi - \theta_x, \theta_y, 0)$ are locally equivalent to each other.
This degeneracy is indicated by the colored base in Fig.~\ref{fig:schematic}~(e).
After obtaining $\theta_i$ within the Weyl chamber, we reconstruct the nonlocal gate from Eq.~\eqref{eq:two_qubit_gate}.

We compare the time-evolution $U_\mathcal{T}$, obtained by solving the time-dependent Schrödinger equation for Eq.~\eqref{eq:effective_hamiltonian}, with a target gate $U_T$.
We define the fidelity $\mathcal{F}(U_\mathcal{T}, U_T)$ as a measure of the closeness between the gates~\cite{Pedersen2007}
\begin{equation}
\mathcal{F}(U_\mathcal{T}, U_T) = \frac{1}{20}\left[\Tr(M M^\dagger) + \abs{\Tr(M)}^2\right],
	\label{eq:fidelity}
\end{equation}
where $M \equiv \tilde{U}_\mathcal{T}^\dagger \tilde{U}_T$.
The tilde indicates that only the nonlocal component of each gate is considered.

{\bf \emph{Isotropic exchange.--}}
Without SOI and under a homogeneous Zeeman field along $z$, the exchange interaction is isotropic, and yields an analytically solvable 2Q gate.
In this case,
\begin{equation}
U_\mathcal{T} = \begin{pmatrix}
  e^{-i \left[\mathcal{T}\Delta_z / \hbar + \Omega\right]} & 0 & 0 & 0 \\
  0 & \cos\Omega & -i\sin\Omega & 0 \\
  0 & -i\sin\Omega & \cos\Omega & 0 \\
  0 & 0 & 0 & e^{i\left[\mathcal{T}\Delta_z / \hbar - \Omega\right]}
 \end{pmatrix},
\end{equation}
where $\Delta_z = \mu_B g^* B_z$ is the Zeeman splitting, and $\hbar \Omega = \int_0^\mathcal{T} J_0(t) dt/2 = \frac{1}{v}\int_{d_\mathrm{max}}^{d_\mathrm{min}} J_0(x)dx + J_0(d_\mathrm{min}) t_w / 2$ is the integrated exchange interaction.
Using Eq.~\eqref{eq:makhlin_invariants}, the Makhlin invariants are $G_1(U_\mathcal{T}) = e^{-2i\Omega}\left(3 + e^{4i\Omega}\right)^2 / 16$ and $G_2(U_\mathcal{T}) = 3\cos(2\Omega)$.
Tuning $v$ or $t_w$ varies $\Omega$, yielding different 2Q gates.
For instance, $\Omega = \pi / 4 + k \pi$ gives a $\sqrt{\mathrm{SWAP}}$ gate, while $\Omega = \pi / 2 + k \pi$ produces a SWAP gate, with $k \in \mathds{Z}$.
More generally, any gate in the SWAP-like family, $\mathrm{SWAP}^\alpha$, can be reached with $\alpha = 2\Omega / \pi$.
This matches the expected result for isotropic exchange interaction~\cite{Burkard2023}.
However, no combination of $v$ and $t_w$ in this setting produces a $\mathrm{CZ}$ gate ($G_1 = 0$, $G_2 = 1$).

{\bf \emph{Longitudinal gradient.--}}
We now consider isotropic exchange interactions and a $B$ gradient along $z$.
This results in a $\mathrm{CPHASE}(\theta)$ gate, with $\theta$ tunable via $v$ and $t_w$.
The $\mathrm{CZ}$ gate is a special case of $\mathrm{CPHASE}(\theta)$ with $\theta = \pi$.
However, the $\mathrm{SWAP}^\alpha$ family is no longer achievable.
A similar effect arises with an inhomogeneous $g^*$-factor.
Gate fidelities are given in the SM~\cite{SupMat}.

\begin{figure}[t!]
	\centering
	\includegraphics[width=\linewidth]{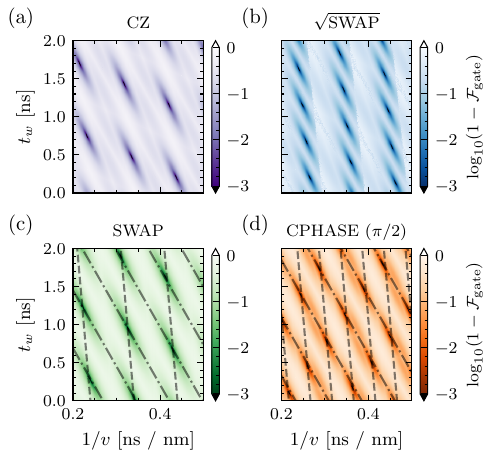}
	\caption{Logarithm of the infidelity $1 - \mathcal{F}$ for a system with $\vec{B} = B \hat{z}$ and $\hat{n}_\mathrm{SOI} = \hat{y}$.
    Target gates are shown on top of each panel.
	In (c, d), we show with dashed and dotted-dashed lines the resonant conditions, see text for details.
    We use $B = 500$~mT, $l_0 = 27.6$~nm, $l_\mathrm{SOI} = 100$~nm.
    }
	\label{fig:infidelities}
\end{figure}

{\bf \emph{Anisotropic exchange.--}}
We now include a moderate SOI, with $l_\mathrm{SOI} \gtrsim l_0$, and constant $\vec{B} \perp \vec{v}_\mathrm{SOI}$, leading to a weakly anisotropic $\overline{J}$.
We later contrast this with systems featuring large SOI.
Remarkably, in this setup, both $\mathrm{CPHASE}(\theta)$ and $\mathrm{SWAP}^\alpha$ families are reachable by appropriately controlling the shuttling parameters.
In Fig.~\ref{fig:infidelities}, we show the infidelity $1 - \mathcal{F}$ as a function of $1 / v$ and $t_w$ for different target gates.
All shown gates reach $\mathcal{F} > 0.999$, neglecting decoherence and other noise sources.
We have used $d_\mathrm{max} = 200$~nm and $d_\mathrm{min} = 80$~nm, yielding $J_0(d_\mathrm{max}) \sim 5\times 10^{-5}$~$\mu$eV, and $J_0(d_\mathrm{min}) \sim 11.5$~$\mu$eV, potentially accessible in current devices~\cite{Weinstein2023}.
We target large $J_0$ to speed up 2Q gates, but results hold for smaller values at the expense of slower gate speed.
In the SM~\cite{SupMat}, we show that this protocol is more robust to common systematic errors than static 2Q gates.

All panels in Fig.~\ref{fig:infidelities} show similar interference patterns of two oscillating functions, with 
\begin{equation}
	\mathcal{F} \sim \left[1 + \sum_{i=1}^2 A_i\cos(k_{x, i} / v_s + k_{y, i} t_w + \phi_i)\right] / 2,
\end{equation}
and $\sum_i A_i = 1$.
Each oscillating function yields repeating resonant lines in the $1 / v$--$t_w$~plane, with slope given by the ratio $k_{y,i}/k_{x,i}$, see Fig.~\ref{fig:infidelities}~(c, d).
The products $k_{x,i}\times k_{y,i}$ and angles $\phi_i$ depend on the target gate.
Qualitatively, we find that the dotted-dashed lines have a slope determined by $\int_0^\mathcal{T} J_0(t)$.
High-fidelity gates emerge when these intersect dashed lines, whose slope depends on $\int_0^\mathcal{T}\alpha(t)$.

\begin{figure}[t!]
	\centering
	\includegraphics[width=\linewidth]{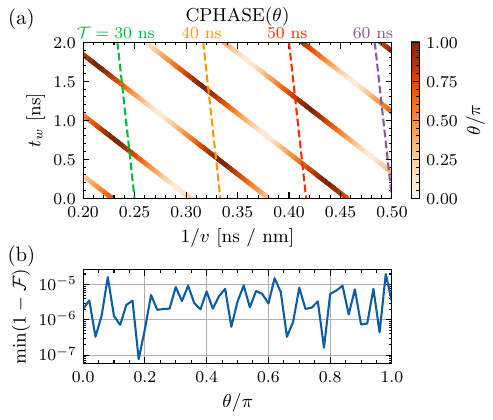}
	\caption{(a) Optimal $t_w$ and $v$ for the implementation of $\mathrm{CPHASE}(\theta)$ gates.
	The color scale indicates the angle $\theta$.
	Colored dashed lines indicate curves of constant gate time $\mathcal{T}$, as shown on the top.
	(b) Minimum infidelity versus $\theta$.
	Our protocol leads to high gate fidelity above $99.998\%$ across all possible $\theta$.
	All parameters are the same as in Fig.~\ref{fig:infidelities}.}
	\label{fig:CPHASE}
\end{figure}

Beyond the examples in Fig.~\ref{fig:infidelities}, other gates are also accessible using the same protocol.
In Fig.~\ref{fig:CPHASE}~(a), we show the optimal $t_w$ and $v$ for implementing $\mathrm{CPHASE}(\theta)$ gates.
Remarkably, all gates lie along straight trajectories in the $1/v$\nobreakdash--$t_w$~plane, facilitating experimental implementation and avoiding the need for complex optimization.
Each replica contains all angles $\theta\in [0,\pi]$, and all replicas have identical fidelity for a given $\theta$.
Fig.~\ref{fig:CPHASE}~(b) shows that the entire class of $\mathrm{CPHASE}(\theta)$ gates is achievable with $\mathcal{F} > 0.99998$ for any $\theta$.

The entangling power~\cite{Zanardi2000}, i.e., the average entanglement a gate generates when acting on product states, for a CPHASE$(\theta)$ gate is given by $e_p(\theta) = (1 - \cos\theta)/9$.
Using moderate SOI, we continuously tune the entangling power from $e_p(0) = 0$ to $e_p(\pi) = 2/9$, corresponding to a fully entangling 2Q gate.
Control over the entangling power is crucial for quantum algorithms, including quantum variational circuits~\cite{Nakhl2024}.
We estimate that, in state-of-the-art experiments, CPHASE$(\pi/2)$ is reached in $\mathcal{T} \sim 25$~ns and CPHASE$(\pi)$ in $\mathcal{T} \sim 28$~ns.
Faster gates are possible by increasing the shuttling speed.
Moreover, increasing $J_0(d_\mathrm{min})$ reduces the slope of the high-fidelity lines until they align with constant gate-time contours.
This enables any $\mathrm{CPHASE}(\theta)$ to be implemented at fixed gate time, a key feature for quantum error correction, as the accumulated error becomes independent of $\theta$.

\begin{figure}[t!]
	\centering
	\includegraphics[width=\linewidth]{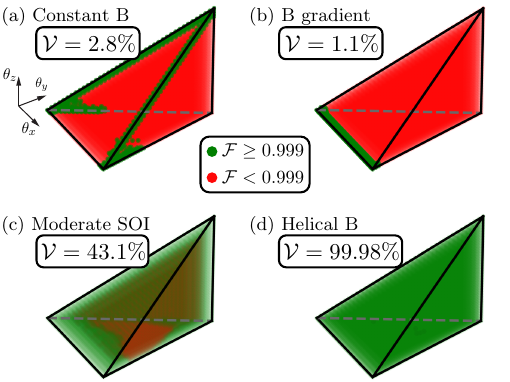}
	\caption{Weyl chamber coverage for different configurations.
    The percentage of target gates with high fidelity is shown in the top left corner of each panel.
	The helical $B$ field with our shuttling protocol enables an almost arbitrary 2Q gate realization by a single step.}
	\label{fig:weyl_chambers}
\end{figure}

{\bf \emph{Weyl chamber.--}}
To benchmark the coverage of 2Q gates enabled by our shuttling protocol, we divide the Weyl chamber into a grid of equidistant points and compute the fraction ($\mathcal{V}$) of gates reachable with $\mathcal{F} > 0.999$.
We define the coverage as the percentage of 2Q gates that can be reached with high fidelity.
Low coverage indicates that accessible gates are confined to a small subset of the chamber, while high coverage implies access to most of it, enabling an arbitrary 2Q gate in a single step.
In Fig.~\ref{fig:weyl_chambers}, we present this coverage for various system configurations.
We consider shuttling speeds $v\in [0.01, 10]$~m/s and waiting times $t_w \in [0, 10]$~ns.
For a constant $B$ and no SOI, the coverage is $2.8\%$, limited to the edges connecting $(0,0,0)$, $(\pi/2,\pi/2,\pi/2)$, and $(\pi,0,0)$, corresponding to $\mathrm{SWAP}^\alpha$ gates.
Including an inhomogeneous magnetic field $\vec{B}(x) = (B_0 + \Delta_B x)\hat{z}$ with $B_0 = 0.5$~T and $\Delta_B = 0.25$~mT/nm reduces the coverage to $1.1\%$, allowing only $\mathrm{CPHASE}(\theta)$ gates.

By contrast, $l_\mathrm{SOI}=100$~nm yields a coverage of $43.1\%$, enabling high fidelities for all edges of the Weyl chamber and all faces except the base $(\theta_x, \theta_y, 0)$.
Besides $\mathrm{SWAP}^\alpha$ and $\mathrm{CPHASE}(\theta)$ gates, this configuration also allows direct implementation of the fermionic simulation gate~\cite{Tsoukalas2025}, fSim$(\theta, \phi)$, a native gate in Google's processors~\cite{Arute2019,Foxen2020}, defined as
\begin{equation}
	\begin{split}
		U_\mathrm{fSim}(\theta, \phi) = &\exp\left\{-i\left[\theta\left(\sigma_x\otimes \sigma_x + \sigma_y\otimes \sigma_y\right) / 2 \vphantom{\sigma_z^{(1)}}\right.\right.\\
		& \left.\left. + \phi\left(\mathds{I}-\sigma_z^{(1)}-\sigma_z^{(2)}+\sigma_z\otimes \sigma_z\right)\right]\right\}.
	\end{split}
\end{equation}
With SOI, we achieve $\mathcal{F} > 0.999$ for general fSim$(\theta, \phi)$ gates.
However, the bulk of the Weyl chamber remains inaccessible.
This region corresponds to gates with $\theta_x \neq \theta_y$ in Eq.~\eqref{eq:two_qubit_gate}.
An example is the Berkeley gate, a maximally entangling operation defined as $\mathcal{B}=\exp[i\pi/2(2\sigma_x\otimes \sigma_x+\sigma_y\otimes\sigma_y)]$.
This gate enables the synthesis of any 2Q operation with a minimal number of single- and two-qubit gates~\cite{Zhang2004}, outperforming CNOT and double-CNOT constructions.

\begin{table*}[t!]
    \centering
    \begin{tabular}{@{}lcccccc@{}}
        \toprule
        \textbf{Configuration} & \textbf{CPHASE}($\theta$) & $\mathbf{SWAP}^\alpha$ & \textbf{iSWAP} &  \textbf{fSim}($\theta, \phi$) &  $\boldsymbol{U_{2Q}}(\theta_x, \theta_y, \theta_z)$\\
        \midrule
        Constant $B$      & \xmark & \cmark & \xmark & \xmark & \xmark \\
		$B$ gradient      & \cmark & \xmark & \xmark & \xmark & \xmark \\
		Moderate SOI      & \cmark & \cmark & \cmark & \cmark & \xmark \\
		Helical $B$       & \cmark & \cmark & \cmark & \cmark & \cmark \\
        \bottomrule
    \end{tabular}
	\caption{Accessibility of 2Q gates for different system configurations.
	For parametrized gates, e.g., CPHASE($\theta$), we only mark it as accessible if the full range of parameter(s) is reachable with high fidelity $\mathcal{F} > 0.999$.}
    \label{tab:gates_vs_config}
\end{table*}

{\bf \emph{Large SOI-enabled arbitrary 2Q gate.--}}
To implement any possible 2Q gate directly, we consider systems with a helical Zeeman field induced by nanomagnets in Si electron devices.
We explicitly take $\vec{B}(x) = B_0 \left[\sin(2\pi x / \lambda_B) \hat{x} + \cos(2\pi x / \lambda_B) \hat{z}\right]$, rotating in the $x$\nobreakdash--$z$~plane.
Here, $B_0$ is the field amplitude, and $\lambda_B$ the nanomagnet period~\cite{Aldeghi2023}.
Similar physics emerges in hole systems, e.g., in Ge or Si, through SOI and periodic gate-induced $\overline{g}$ modulation, or via combined SOI and $\overline{g}$ engineering~\cite{Bosco_2021,Liles_2021,AbadilloUriel_2023}.
A strong SOI alone also leads to the same effective Hamiltonian.
In that case, the effective Zeeman field~\cite{Bosco2024}: ${E}_z(x) = \exp(-l_0^2 / l_\mathrm{SOI}^2)g^* \mu_B B_0\left[\sin(2x/l_\mathrm{SOI})\hat{x} + \cos(2x/l_\mathrm{SOI})\hat{z}\right]$.

This configuration enables the implementation of almost any 2Q gate in a single step, as demonstrated by the $99.98\%$ Weyl chamber coverage in Fig.~\ref{fig:weyl_chambers}~(d).
Owing to the finite fidelity threshold, nearly the full 3D volume is reachable by tuning just two parameters.
Perfect entanglers lie along the line from $(\theta_x, \theta_y,\theta_z) = (\pi/2, 0, 0)$ to $(\pi/2, \pi/2, 0)$~\cite{Balakrishnan2011,Jonnadula2020}, which becomes fully accessible with helical $B$.
The simulation parameters are $B_0 = 50$~mT and $\lambda_B = 50$~nm.
Comparable periods have been achieved in recent nanomagnet experiments~\cite{Aldeghi2023}.
Achieving the same coverage using only static fields and SOI requires $B_0 \sim 0.9$~T and $l_\mathrm{SOI} = 16\,\mathrm{nm} < 27.6\, \mathrm{nm} = l_0$.
Such SOI lengths have been reported in Si FinFETs~\cite{Geyer2024} and Ge/Si nanowire QDs~\cite{Froning_2021}.
Comparable SOI lengths can be achieved in planar Ge via strain engineering~\cite{AbadilloUriel_2023,Costa_2025,Mauro_2025} or squeezed QDs~\cite{Bosco_2021}.

In Table~\ref{tab:gates_vs_config}, we summarize the accessibility of different 2Q gates for the studied configurations.
Other variants, such as a helical SOI, are detailed in the SM~\cite{SupMat}.
Large SOI and helical $B$ also mitigate noise and improve gate fidelity by filtering low-frequency noise.
This behavior is consistent with previous observations in single-qubit systems~\cite{Bosco2024}.
We confirm that the same mechanism enhances fidelity in our 2Q system and provide a detailed analysis in the SM~\cite{SupMat}.
Finally, in the SM~\cite{SupMat}, we simulate a state-of-the-art Si device with two micromagnets and show that the Weyl chamber coverage reaches $86\%$, underscoring the viability of our protocol in current devices and its near-term experimental implementation.

Achieving full Weyl chamber coverage is crucial for implementing arbitrary 2Q gates in a single step.
Other locally equivalent gates can be selected by applying single-qubit operations, achievable by controlling the velocity before and after the 2Q gate.
Since SOI enables full control of individual spin states during motion, this approach speeds up the full gate sequence~\cite{FernandezFernandez2024}.
Further optimization of gate time and robustness against noise sources, such as charge noise, can be achieved using pulse-engineering techniques~\cite{Ban2018,Ban2019,FernandezFernandez2022,Meinersen2024,Meinersen2025}, including shortcuts to adiabaticity~\cite{Chen2010,GueryOdelin2019} or reinforcement-learning-based strategies~\cite{Porotti2019,Ding2021}.

{\bf \emph{Conclusion.--}}
We studied 2Q gates performed simultaneously with spin shuttling in a semiconductor double QD using a conveyor-mode potential.
We developed a general formalism to compute the effective Hamiltonian in the presence of position-dependent magnetic fields, SOI, and anisotropic $g$-tensors.
Our method applies to both electrons and holes and is compatible with various semiconductor platforms, including Si, Ge, GaAs, and InAs, in planar and nanowire QD geometries.

A central result is that large SOI, often viewed as detrimental to coherent spin control, can instead be exploited to expand the accessible set of 2Q gates during shuttling.
By using either helical magnetic fields or intrinsic strong SOI, nearly any 2Q gate can be realized by tuning the shuttling speed and waiting time.
Crucially, all ingredients of the proposed protocol, gate-defined QDs, tunable electrostatic potentials, and spatially varying magnetic fields, are already available or under active development in current experiments.

Our results provide a realistic and immediately implementable route to arbitrary 2Q gates in scalable semiconductor devices.
This approach circumvents the need for complex sequences or long-range static couplings, enabling direct 2Q operations between initially separated spins, a key advantage for distributed architectures, and fault-tolerant quantum computation.

\emph{Acknowledgments.--}
We thank the members of the Rimbach-Russ and Bosco group for useful discussion.
G.P. and D.F.F. are supported by the Spanish Ministry of Science through the grant: PID2023-149072NB-I00 and by the CSIC Research Platform PTI-001.
D.F.F. acknowledges support from FPU Program No. FPU20/04762.
Y.M., L.M.K.V., and S.B. acknowledge support by the EU through the H2024 QLSI2 project, the Army Research Office under Award Number: W911NF-23-1-0110, and NCCR Spin Grant No. 51NF40-180604.
The views and conclusions contained in this document are those of the authors and should not be interpreted as representing the official policies, either expressed or implied, of the Army Research Office or the U.S. Government.
The U.S. Government is authorized to reproduce and distribute reprints for Government purposes notwithstanding any copyright notation herein.

\newpage

\begin{widetext}

\begin{center}
\textbf{\large Supplemental Material: Spin-orbit-enabled realization of arbitrary two-qubit gates on moving spins \\}
\end{center}
\setcounter{section}{0}
\setcounter{equation}{0}
\setcounter{figure}{0}
\setcounter{table}{0}

\setcounter{page}{1}
\renewcommand{\theequation}{S\arabic{equation}}
\renewcommand{\thefigure}{S\arabic{figure}}
\renewcommand{\thetable}{S\arabic{table}}


\section{Numerical details} \label{sec:numerical_details}
To obtain the effective Hamiltonian, we numerically diagonalize the total Hamiltonian given in Eq.~(1) in the main text.
We discretize the Hamiltonian using a finite difference method on a grid along the $x$ axis for both particles.
The grid is defined as $\vec{x} = (x_i, x_j) = (i\cdot \Delta x, j \cdot \Delta x)$, with $i,j \in [0, N_x]$ and spacing $\Delta x = 4$~nm.
We use $N_x = 101$ points and verify that refining the grid does not affect the results.

Operators are discretized using second-order accurate central finite differences.
Special care is taken with the spin-orbit interaction (SOI) term when the SOI vector is position dependent
\begin{equation}
	\begin{split}
	H_\mathrm{SOI} \psi_{i,j} &= -\sum_{k=1}^{2}\left\{\vec{v}_\mathrm{SOI}(x_k), -i\hbar\partial_{x_k} \right\} \cdot \vec{\sigma}_k \psi_{i,j} \\
	&= \frac{-i\hbar}{4\Delta x} \left[(\vec{v}_{\mathrm{SOI}, i} + \vec{v}_{\mathrm{SOI}, i + 1})\cdot \vec{\sigma}_1 \psi_{i+1, j} - (\vec{v}_{\mathrm{SOI}, i} + \vec{v}_{\mathrm{SOI}, i - 1})\cdot \vec{\sigma}_1 \psi_{i-1, j} + (1\leftrightarrow 2, i \leftrightarrow j)\right],
	\end{split}
	\label{eq:discretized_SOI}
\end{equation}
where $\{A, B\}= (AB + BA)/2$ is the normalized anticommutator, and $\psi_{i,j}$ is the discretized spinor with dimension $2\times 2$ corresponding to the spin degrees of freedom.
The same care must be taken when numerically solving the Schrödinger equation in presence of inhomogeneous mass \cite{Harrison2016}.

For a general operator acting on a single particle, the finite difference method reads
\begin{equation}
	\hat{O} \psi_{i,j} = a_i \psi_{i-1, j} + b_i \psi_{i, j} + c_i \psi_{i+1, j}.
\end{equation}
Here, we have assumed that the operator $\hat{O}$ contains only first and second derivatives, such as is the case of the kinetic and the SOI terms.
To ensure that the operator is Hermitian, we must have $a_{i + 1} = c_i^*$.
Looking at Eq.~\eqref{eq:discretized_SOI}, we can see that the SOI term is Hermitian.

Finally, we include a Coulomb interaction term
\begin{equation}
	H_\mathrm{C} = \frac{1}{4\pi \epsilon}\frac{e^2}{\abs{x_1 - x_2}}.
\end{equation}
For $x_1 = x_2$ we have a singularity, and we must use a regularization procedure.
We use a regularization of the form
\begin{equation}
	\abs{x_1 - x_2} \sim \sqrt{\abs{x_1 - x_2}^2 + \delta^2},
\end{equation}
where $\delta$ is a small regularization parameter, preventing the singularity.
This softening of the Coulomb interaction is sometimes called soft-core Coulomb interaction \cite{Gebremedhin2014,Inarrea2019,Truong2022}.
In our simulations we use $\delta = 20$~nm.
The minimum distance between the two particles that we consider is  $d_\mathrm{min} = 80\; \mathrm{nm} \gg \delta$, and not visible deviation from the Coulomb interaction regularization is expected.

\section{Effective Hamiltonian} \label{sec:effective_hamiltonian}
The effective Hamiltonian is obtained by projecting the full Hamiltonian given in Eq.~(1) in the main text onto the computational basis $\{\ket{\uparrow,\uparrow}, \ket{\uparrow,\downarrow}, \ket{\downarrow,\uparrow}, \ket{\downarrow,\downarrow}\}$.
First, we compute the eigenstates of the discretized Hamiltonian using the Implicitly Restarted Lanczos Method.
We compute up to the first 200 lowest energy eigenstates.
However, since we are working with identical fermions, we ensure a proper antisymmetrization of the eigenstates.
For that, we compute
\begin{equation}
	\phi^A_k(x_1, x_2, s_1, s_2) = \frac{1}{\sqrt{2}}\left[\phi(x_1, x_2, s_1, s_2) - \phi(x_2, x_1, s_2, s_1)\right],
\end{equation}
where $\phi_k(x_1, x_2, s_1, s_2)$ is the $k$-th eigenstate of the discretized Hamiltonian, and $s_i$ is the spin degree of freedom of the $i$-th particle.
Sometimes, the diagonalization algorithm results in two eigenstates that after antisymmetrization are equal.
In this case, we compute the overlap between all pairs of eigenstates with equal eigenenergy.
If the overlap is equal to one, up to machine precision, we assume that the two eigenstates are equal, and we remove one of them.

To construct the computational basis, we employ a Hund-Mulliken molecular orbital theory.
We start from the single particle wave functions given by the diagonalization of the single particle Hamiltonian, which reads as
\begin{equation}
	H_\mathrm{L (R)} = H_\mathrm{K, L(R)} + V_\mathrm{L(R)}(x, t),
\end{equation}
consisting of the kinetic and the potential energy terms for the left (L) and right (R) quantum dots.
We approximate the local Hamiltonians by a harmonic oscillator with a confinement frequency $\omega_0$, centered at the position of the left and the right quantum dots $x = \pm d / 2$.
Using the ground states of the harmonic oscillators $\ket{\psi_0^\mathrm{L(R)}}$, we construct the maximally localized Wannier states as
\begin{equation}
	\ket{\mathrm{L(R)}} = \sqrt{N}\left(\ket{\psi_0^\mathrm{L(R)}}-\gamma \ket{\psi_0^\mathrm{R(L)}}\right),
\end{equation}
where $S = \braket{\psi_0^\mathrm{L}}{\psi_0^\mathrm{R}}$ is the overlap between the two single particle wave functions, $\gamma = (1 - \sqrt{1 - S^2})/S$, and $N = (1 - 2 \gamma S+\gamma^2)^{-1}$ is a normalization factor.
With the above definition we ensure that the Wannier states is orthonormal $\braket{\mathrm{L}}{\mathrm{R}} = 0$.
Finally, the computational basis in constructed using Slater determinants, resulting in the following four states
\begin{equation}
	\begin{split}
	\ket{\uparrow,\uparrow} &= \left[\ket{\mathrm{L}_\uparrow(x_1), \mathrm{R}_\uparrow(x_2)} - \ket{\mathrm{L}_\uparrow(x_2), \mathrm{R}_\uparrow(x_1)}\right]/\sqrt{2},\\
	\ket{\uparrow,\downarrow} &= \left[\ket{\mathrm{L}_\uparrow(x_1), \mathrm{R}_\downarrow(x_2)} - \ket{\mathrm{L}_\downarrow(x_2), \mathrm{R}_\uparrow(x_1)}\right]/\sqrt{2},\\
	\ket{\downarrow,\uparrow} &=\left[\ket{\mathrm{L}_\downarrow(x_1), \mathrm{R}_\uparrow(x_2)} - \ket{\mathrm{L}_\uparrow(x_2), \mathrm{R}_\downarrow(x_1)}\right]/\sqrt{2},\\
	\ket{\downarrow,\downarrow} &=\left[\ket{\mathrm{L}_\downarrow(x_1), \mathrm{R}_\downarrow(x_2)} - \ket{\mathrm{L}_\downarrow(x_2), \mathrm{R}_\downarrow(x_1)}\right]/\sqrt{2}.
	\end{split}
	\label{eq:computational_basis}
\end{equation}

We project high-energy eigenstates onto this basis using a Schrieffer-Wolff (SW) transformation.
When the two particles are too close, i.e., $d /2 \sim l_0$, where $l_0$ is the confinement length scale, the exchange interaction is strong and two different resolved quantum dots are not a good approximation.
As a consequence, the SW transformation does not converge.
The radius of convergence of the SW transformation is given by the condition \cite{Bravyi2011} $\rho_c > 1$, with $\rho_c = \epsilon_c / \left[8 (1 + 2\abs{\mathcal{I}_0} / \pi\Delta)\right]$, where $\mathcal{I}_0$ is the difference in energy between the minimum and the maximum energy of the lowest energy manifold, and $\Delta$ is the gap between the low and high energy manifolds.
Finally, $\epsilon_c = \Delta / ||V||$, where $||V||$ is the Frobenius norm of the interaction term between the low and high energy states.
In Fig.~\ref{fig:SW_convergence}~(a), we plot the convergence radius of the SW transformation as a function of the distance between the two particles.
We can see that for distances smaller than $d \sim 45$~nm, the SW transformation does not converge, see Fig.~\ref{fig:SW_convergence}~(b).
To ensure convergence, we use a minimum distance $d_\mathrm{min}=80$~nm, larger than the threshold, and we use an SW transformation up to order 100.

\begin{figure}[t!]
	\centering
	\includegraphics{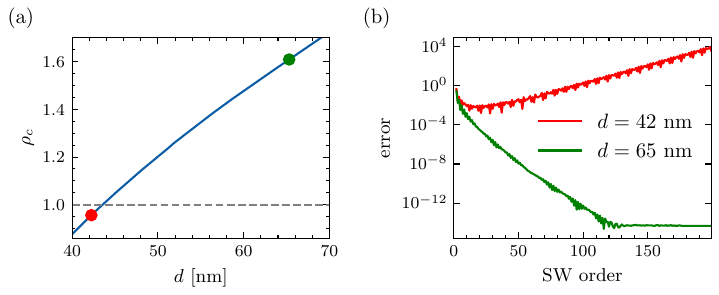}
	\caption{(a) Convergence radius $\rho_c$ versus the distance between the two particles $d$.	
	With a dashed line we show the convergence threshold of $\rho_c = 1$.
	(b) Energy mismatch between the four lowest eigenenergies of the total Hamiltonian and the eigenenergies of the effective Hamiltonian as a function of the SW order, for two different distances $d = 42$~nm (red, diverge) and $d = 65$~nm (green, converge).
	The confinement length scale is $l_0 = 27.6$~nm.
	All other parameters are the same as those used in the main text.
	}
	\label{fig:SW_convergence}
\end{figure}

The effective Hamiltonian is then given by
\begin{equation}
	\tilde{H} = \frac{1}{2}\left(\vec{\Delta}_L \cdot \vec{\sigma}^{(L)} + \vec{\Delta}_R \cdot \vec{\sigma}^{(R)}\right) + \frac{1}{4} \vec{\sigma}^{(L)} \cdot \overline{J} \vec{\sigma}^{(R)}.
\end{equation}
From the numerical effective Hamiltonian, we extract the anisotropic exchange matrix $\overline{J}$.
The interaction term can in general be written as
\begin{equation}
	\vec{\sigma}^{(L)} \cdot \overline{J} \vec{\sigma}^{(R)} = J \vec{\sigma}^{(L)} \cdot \vec{\sigma}^{(R)} + \vec{D}\cdot \vec{\sigma}^{(L)} \times \vec{\sigma}^{(R)} + \vec{\sigma}^{(L)} \overline{\Gamma} \vec{\sigma}^{(R)},
\end{equation}
where $J$ is the isotropic exchange interaction, $\vec{D}$ is the Dzyaloshinskii-Moriya vector, and $\overline{\Gamma}$ is the magnetic anisotropy, a symmetric traceless tensor.
However, in this work we opt to use the more compact expression with the anisotropic exchange interaction matrix $\overline{J}$ written in terms of a rotation matrix in the form $\overline{J} = J_0 R(\theta, \phi, \alpha)$.
The rotation matrix is parametrized by the rotation vector $\hat{n}_R = (\sin\theta \cos\phi, \sin\theta \sin\phi, \cos\theta)$ and the rotation angle $\alpha$.
Both representations are connected by the relations
\begin{equation}
	\begin{split}
		J &= J_0 \cos\alpha, \\
		\vec{D} &= -J_0 \sin\alpha \; \vec{n}_R, \\
		\overline{\Gamma} &= J_0(1 - \cos\alpha) \; \vec{n}_R^T \times \vec{n}_R.
	\end{split}
\end{equation}

Since the eigenvalues of the rotation matrix are complex values with unitary norm, we obtain the exchange interaction as the absolute value of the eigenvalues of $\overline{J}$, and $R = \overline{J} / J_0$.
Secondly, we compute the trace of the rotation matrix, which is bounded by $-1 \leq \Tr(R) \leq 3$.
If the trace is $\Tr(R) = 3$, the rotation matrix is the identity, and the angles are $\theta = \phi = \alpha = 0$.
Otherwise, we can compute the rotation angle $\alpha$ as
\begin{equation}
	\alpha = \arccos\left(\frac{\Tr(R) - 1}{2}\right).
\end{equation}
To compute the polar and azimuthal angles, we first compute the component of the rotation vector, which are given by
\begin{equation}
	\hat{n}_R = \sqrt{\frac{1 + R_{xx}}{2}}\hat{x} + \sqrt{\frac{1 + R_{yy}}{2}}\hat{y} + \sqrt{\frac{1 + R_{zz}}{2}}\hat{z},
\end{equation}
if $\Tr(R) = -1$, and by 
\begin{equation}
	\hat{n}_R = \frac{1}{\sqrt{\left[3 - \Tr(R)\right]\left[1+\Tr(R)\right]}}\left[(R_{zy}-R_{yz})\hat{x} + (R_{xz}-R_{zx})\hat{y} + (R_{yx}-R_{xy})\hat{z}\right],
\end{equation}
otherwise.
Finally, the polar and azimuthal angles are given by
\begin{equation}
	\begin{split}
		\theta &= \arccos\left(n_{R,z}\right), \\
		\phi &= \operatorname{sign}(n_{R,y})\arccos\left(\frac{n_{R,x}}{\sqrt{n_{R,x}^2 + n_{R,y}^2}}\right).
	\end{split}
\end{equation}

By close inspection of the effective Hamiltonian with a general SOI vector given by the spherical angles $\theta_\mathrm{SOI}$ and $\phi_\mathrm{SOI}$, we obtain the relation
\begin{equation}
	\begin{split}
		\theta =& \theta_\mathrm{SOI}, \\
		\phi =& \phi_\mathrm{SOI} + \pi. \\
	\end{split}
\end{equation}

In Fig.~\ref{fig:effective_exchange}, we show the effective exchange interaction as a function of the distance between the two particles $d$ and the SOI length $l_\mathrm{SOI}$.
Here we can see that the effective exchange interaction is independent of the SOI length, and it is an exponential function of the distance between the two particles of the form $J_0 = 1.9 e^{-1.8d/l_0}$~meV, with $l_0 \sim 27.6$~nm being the natural length scale of the confinement.

\begin{figure}[t!]
	\centering
	\includegraphics{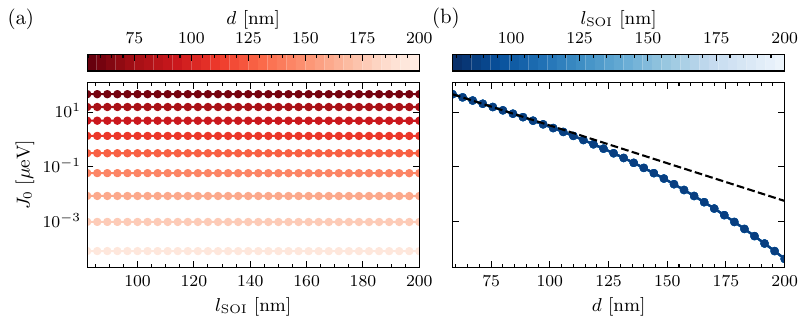}
	\caption{Effective exchange interaction for different values of the distances between the two dots $d$, and the SOI length $l_\mathrm{SOI}$. 
		Other parameters are $\hbar \omega = 1$~meV, $B_z = 50$~mT, $\hat{n}_\mathrm{SOI} = \hat{z}$.
		In panel (b), all values fall in the same region as the blue dots, and are not visible.
		The black dashed line is a fit to an exponential decay.
		}
	\label{fig:effective_exchange}
\end{figure}

Finally, we have to model the effective rotation angle $\alpha$.
We find that a good agreement with the numerical results is obtained by using the following form
\begin{equation}
	\alpha = \frac{a_0 + a_1 \cdot d + a_2 \cdot d^2}{l_\mathrm{SOI}},
\end{equation}
where $a_0$, $a_1$, and $a_2$ are constants that depend on the parameters of the system.
The above functional form closely fits the numerical data, as seen Fig.~\ref{fig:effective_angle}.
These results are obtained for a SOI vector in the $z$ direction, but the same function form for both the exchange interaction and the rotation angle is obtained for other SOI directions.
The numerical results are in good agreement with the analytical results obtained in Ref.~\cite{Qi2024}.

\begin{figure}[t!]
	\centering
	\includegraphics{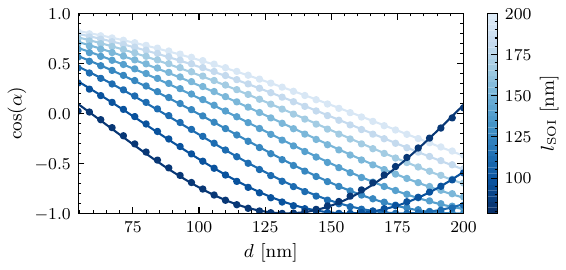}
	\caption{Effective rotation angle for different values of the distances between the two dots $d$, and the SOI length $l_\mathrm{SOI}$.
		All other parameters are the same as in Fig.~\ref{fig:effective_exchange}.
		Dots are numerical data, and colored lines are fits to the function $\cos(\alpha) = \cos\left[(a_0 + a_1 \cdot d + a_2 \cdot d^2) / l_\mathrm{SOI}\right]$.
		}
	\label{fig:effective_angle}
\end{figure}

\section{Additional system configurations} \label{sec:additional_configurations}
In this appendix, we present the fidelity for the gate set $\mathrm{CNOT}$, $\sqrt{\mathrm{SWAP}}$, $\mathrm{SWAP}$, and $\mathrm{CPHASE}(\pi /2)$ for the configurations presented in the main text, as well as for additional configurations.

\subsection{Constant magnetic field}
In Fig.~\ref{fig:infidelities_benchmark_UT}, we show the fidelities for different two-qubit (2Q) gates versus the shuttling velocity and the waiting time, for the case of no SOI, and a constant magnetic field.
In this case, the only accessible gates from the set mentioned above are the $\sqrt{\mathrm{SWAP}}$ and the $\mathrm{SWAP}$ gates.
Here, the lines of high fidelity are in perfect agreement with the analytical result given by
\begin{equation}
	t_w = \frac{2\hbar\Omega}{J_0(d_\mathrm{min})} - \frac{2}{v}\int_{d_\mathrm{max}}^{d_\mathrm{min}} \frac{J_0(x)}{J_0(d_\mathrm{min})}dx,
	\label{eq:waiting_time_benchmark}
\end{equation}
with $\Omega = \pi (1 / 4 + k)$ and $\Omega = \pi (1 / 2 + k)$ for the $\sqrt{\mathrm{SWAP}}$ and $\mathrm{SWAP}$ gates, respectively, with $k \in \mathbb{Z}$.

\begin{figure}[t!]
	\centering
	\includegraphics{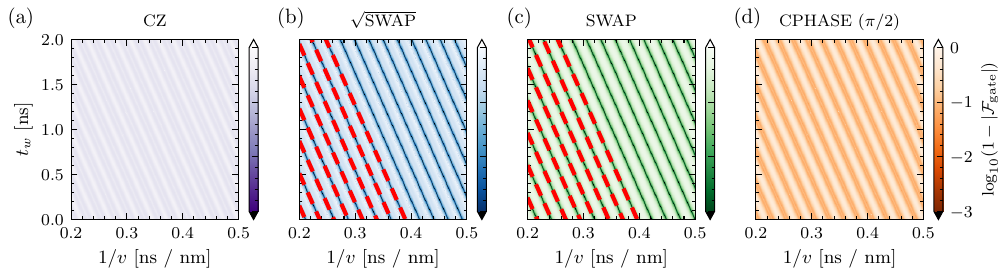}
	\caption{Fidelities for different 2Q gates versus the shuttling velocity and the waiting time, for the case of no SOI, and constant magnetic field.
	The parameters are $\omega = 1$~meV, $\vec{B} = 0.5 \hat{z}$~T, $d_{\min} = 80$~nm, and $d_{\max} = 200$~nm.
	The red dashed lines in panels (b) and (c) denote the analytical prediction for the waiting times $t_w$ for the $\sqrt{\mathrm{SWAP}}$ and $\mathrm{SWAP}$ gates, respectively, given by Eq.~\eqref{eq:waiting_time_benchmark}.
	}
	\label{fig:infidelities_benchmark_UT}
\end{figure}

\subsection{Magnetic field gradient}
When adding a longitudinal magnetic field gradient as $\vec{B} = (B_0 + \Delta_B x)\hat{z}$, the $\mathrm{CNOT}$ and $\mathrm{CPHASE}(\pi /2)$ gates become accessible but the $\sqrt{\mathrm{SWAP}}$ and $\mathrm{SWAP}$ gates are out of reach, see Fig.~\ref{fig:infidelities_magnetic_field_gradient_z_direction_UT}.
\begin{figure}[t!]
	\centering
	\includegraphics{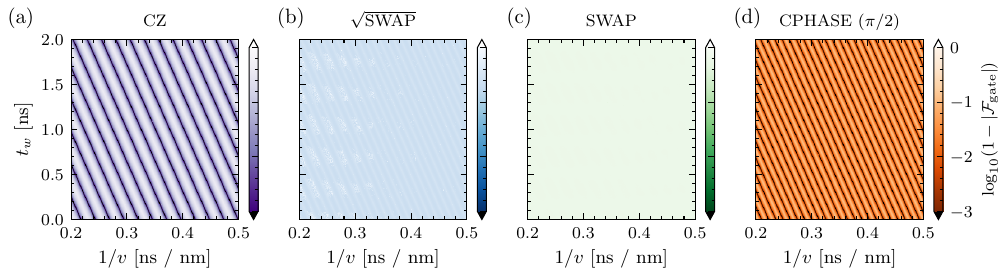}
	\caption{Fidelities for different 2Q gates versus the shuttling velocity and the waiting time, for the case of no SOI, and magnetic field gradient.
	The parameters are $\omega = 1$~meV, $B_0 = 0.5$~T, $\Delta B = 0.25$~mT/nm, $d_{\min} = 80$~nm, and $d_{\max} = 200$~nm.
	}
	\label{fig:infidelities_magnetic_field_gradient_z_direction_UT}
\end{figure}

\subsection{Helical magnetic field}
In Fig.~\ref{fig:infidelities_rotating_magnetic_field_4_UT}, we show the fidelities for  case of a helical magnetic field.
Here, we can see that due to the fast helical magnetic field, the interference patter is more rapidly oscillating.
For slow shuttling velocities $1 / v > 2$~ns / nm, the dynamics is adiabatic, and a repeating pattern is observed.
However, for fast shuttling velocities $1 / v < 2$~ns / nm, the dynamics is non-adiabatic, and the interference pattern is more complex.
It is the fast shuttling regime which allows for a full coverage of the Weyl chamber.
In the lower panels of Fig.~\ref{fig:infidelities_rotating_magnetic_field_4_UT}, we show a zoom-in in the non-adiabatic regime.

\begin{figure}[t!]
	\centering
	\includegraphics{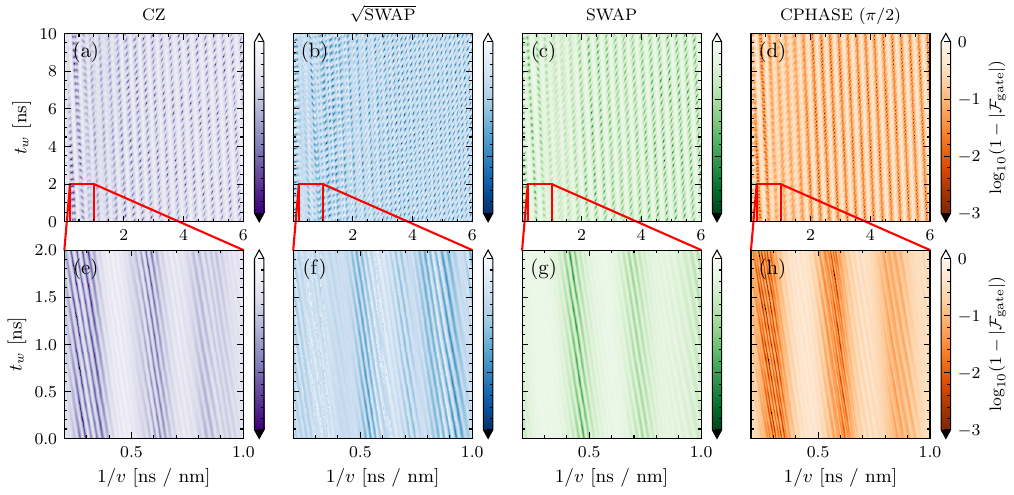}
	\caption{Fidelities for different 2Q gates versus the shuttling velocity and the waiting time, for the case of no SOI, and helical magnetic field.
	The lower panels are a zoom-in of the upper panels in the highlighted region.
	The parameters are $\omega = 1$~meV, $B_0 = 500$~mT, $\lambda_B=50$~nm, $d_{\min} = 80$~nm, and $d_{\max} = 200$~nm.
	}
	\label{fig:infidelities_rotating_magnetic_field_4_UT}
\end{figure}

\subsection{Helical spin-orbit interaction}
Finally, we study the case of a helical SOI as:
\begin{equation}
	\vec{v}_\mathrm{SOI}(x)=\frac{v_0}{\sqrt{1 + A^2}}\left[\hat{x}-A \sin \left(\frac{2 x}{\lambda_N}\right) \hat{y}+A \cos \left(\frac{2 x}{\lambda_N}\right) \hat{z}\right],
\end{equation}

This configuration can be achieved by a highly inhomogeneous potential landscape along the $x$ direction, such that the SOI vector oscillates in time during the shuttling.
The results are similar to the case of a static SOI, see Fig.~\ref{fig:infidelities_nutating_spin_orbit_UT}.
The coverage of the Weyl chamber is also close to the case studied in the main text, with $\mathcal{V}\sim 42\%$.

\begin{figure}[t!]
	\centering
	\includegraphics{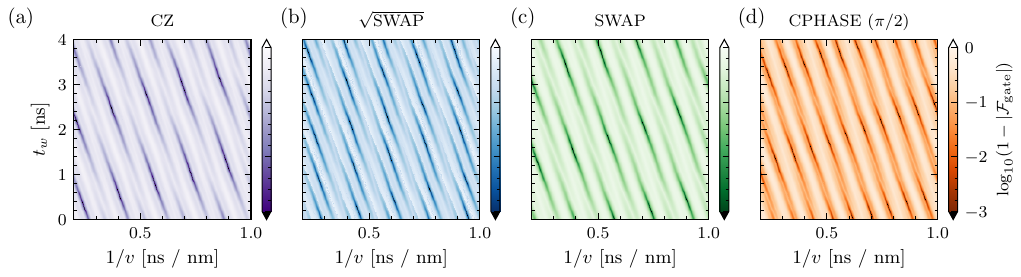}
	\caption{Fidelities for different 2Q gates versus the shuttling velocity and the waiting time, for the case of a helical SOI.
	The parameters are $\omega = 1$~meV, $\lambda_\mathrm{SOI}=\lambda_N= 100$~nm, $A = 0.2$, $B_z = 100$~mT, $d_{\min} = 80$~nm, and $d_{\max} = 200$~nm.}
	\label{fig:infidelities_nutating_spin_orbit_UT}
\end{figure}

\subsection{State-of-the-art silicon spin qubit devices}
Complex magnetic field configurations, such as the helical magnetic field, could become available in the near future, but are not yet available in most spin qubit platforms, including silicon spin qubits.
For this reason, we study the case of a silicon state-of-the-art device, with a maximum exchange interaction $J_0(d_\mathrm{min}) \sim 0.8\; \mu$eV, which is higher than most present experiments, although it should not be out of reach.
This value is obtained in our simulations by fictitiously increasing the Coulomb interaction strength by a factor of $2$, but we stress that similar results are obtained by modifying the confinement potential.
In particular, we reduce the exchange interaction by reducing the dielectric constant of the substrate in our simulations.

For the position dependent magnetic field, we simulate the field generated by a micromagnet close to the central region of the shuttling path, identical to current silicon spin qubit devices \cite{Philips2022, DeSmet2025}.
The magnets are uniformly magnetized along the y-axis with a vector $M=(0, 250, 0)$~mT.
A homogeneous external field, $B_\mathrm{ext}=(0,-25,0)$~mT, is applied antiparallel to the magnetization.
We assume the remanent magnetization is tunable, for instance, by leveraging magnetic hysteresis~\cite{Unseld2025}.
The resulting magnetic field is shown in Fig.~\ref{fig:single_micromagnet_field}~(a-c).

The center of the micromagnets is placed at $\overline{y} \equiv 77 \;\mu$m.
We aim to maximize the coverage of the Weyl chamber by shifting the shuttling path slightly off-center in the $y$-direction.
For that, we study the coverage when placing the shuttling path at $y = y_0$, as shown in Fig.~\ref{fig:single_micromagnet_field}~(d).
We find two maxima, one at $y_0 - \overline{y} \sim 4$~nm and another at $y_0 - \overline{y} \sim 26$~nm, independent of the fidelity threshold.
When setting $\mathcal{F}_\mathrm{th.}=0.999$, we obtain a Weyl chamber coverage of $\mathcal{V} \sim 86\%$ and $\mathcal{V} \sim 40\%$, for the first and second maximum, respectively.
If we lower the fidelity threshold to $\mathcal{F}_\mathrm{th.}=0.995$, we obtain a Weyl chamber coverage of $\mathcal{V} \sim 100\%$ for the first  resonance point, allowing the implementation of almost all 2Q gates with high fidelity.
Let us focus on the first resonance point, which allows for the implementation of a large set of 2Q gates.

\begin{figure}[h!]
	\centering
	\includegraphics{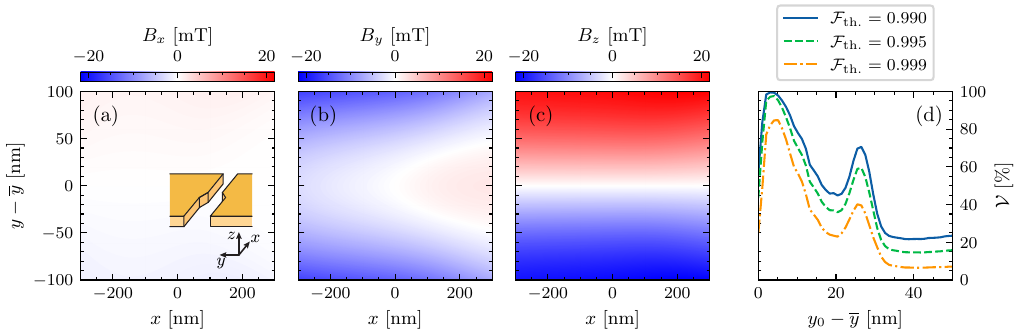}
	\caption{(a-c) Magnetic field generated by a micromagnet.
	The disposition of the micromagnet is shown in the inset of panel (a).
	(d) Weyl chamber coverage versus the position of the shuttling path $y_0$ for different fidelity thresholds.
	}
	\label{fig:single_micromagnet_field}
\end{figure}

At $y_0 - \overline{y} \sim 4$~nm, the micromagnet generates a magnetic field in the $z$-direction that is unchanged as the electron moves along the $x$-axis, while there is a gradient in the $y$-component, as shown Fig.~\ref{fig:simulated_micromagnet_modified}~(a).
With this configuration, we can perform all gates in the $\mathrm{SWAP}$ family, in the diabatic regime, for $1 / v < 1\; \mathrm{ns/nm}$.
Furthermore, working in the adiabatic regime, we can perform other gates with high-fidelity such as the $\mathrm{CPHASE}(\theta=\pi/2)$, see Fig.~\ref{fig:simulated_micromagnet_modified}~(b).
Interestingly, we can also achieve the Berkeley gate, which is usually not accessible, as seen in the main text.
Furthermore, this configuration allows for high-fidelity gates located in the bulk of the Weyl chamber, as shown in Fig.~\ref{fig:simulated_micromagnet_modified}~(c).
As mentioned before, the total Weyl chamber coverage is $\mathcal{V}=86\%$, representing a significant improvement over configurations with a constant magnetic field, or a magnetic field gradient in the $z$-direction, as discussed in the main text.

\begin{figure}[h!]
	\centering
	\includegraphics{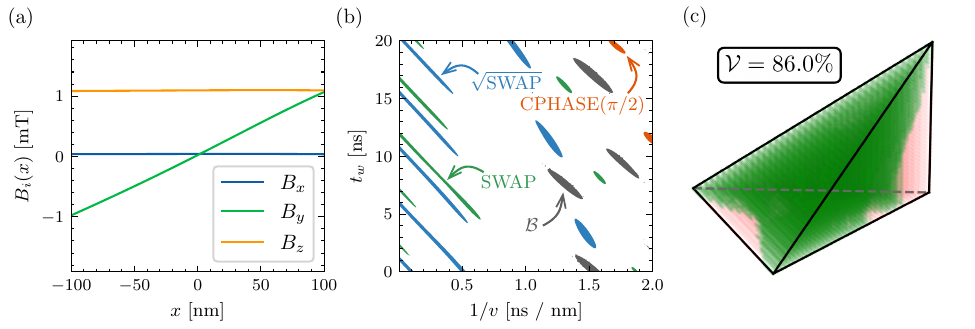}
	\caption{(a) Magnetic field generated by a micromagnet placed, with the shuttling path shifted to $y_0 - \overline{y} = 4$~nm.
	(b) Contours areas with high fidelity ($\mathcal{F} > 0.99$) 2Q gates, color-coded, and marked by the colored arrows.
	(c) Coverage of the Weyl chamber, with green dots indicating 2Q gates with high fidelity ($\mathcal{F}_\mathrm{th.} = 0.999$).
	}
	\label{fig:simulated_micromagnet_modified}
\end{figure}

\clearpage

\section{Systematic errors} \label{sec:systematic_errors}

In this section, we study the effects of a systematic error in the parameters defining the shuttling protocol, i.e., the shuttling speed and the waiting time.
In Fig.~\ref{fig:contour_plots}, we show the contours for areas with fidelity $\mathcal{F} > 0.99$ for different 2Q gates, for each of the configurations studied in the main text.
For the constant magnetic field, and the magnetic field gradients, we can observe thin lines of high fidelity.
When SOI is present, the contours resemble ellipses centered at the optimal points.
The widths of these ellipses is in general larger than the former cases, indicating a more robust protocol against systematic errors.
In the case of a helical magnetic field, the ellipses are more elongated in a given direction defined by the shuttling parameters, but narrower in the perpendicular direction.
Due to the fast oscillations of the two qubits in this case, the contours for some gates are smaller than other gates, indicating that the protocol must be optimized for each gate, specially for the $\mathrm{SWAP}$ gate.

\begin{figure}[h!]
	\centering
	\includegraphics{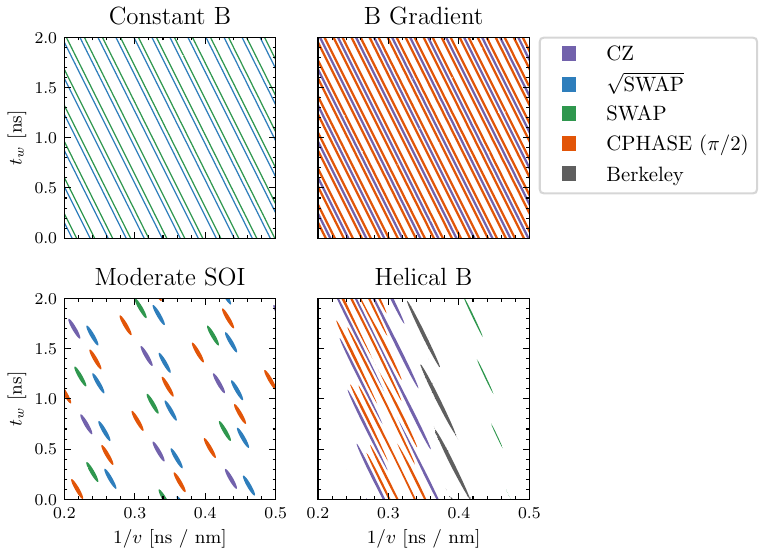}
	\caption{Contours areas with high fidelity ($\mathcal{F} > 0.99$) 2Q gates, color-coded, for different system configurations.
	}
	\label{fig:contour_plots}
\end{figure}

More formally, we study such systematic errors by considering the modified shuttling parameters
\begin{equation}
	v = v^* + \Delta v, \quad t_w = t_w^* + \Delta t_w,
\end{equation}
where $v^*$ and $t_w^*$ are the optimal shuttling speed and waiting time, respectively, and $\Delta v$ and $\Delta t_w$ are the systematic errors in the shuttling speed and the waiting time.
In Fig.~\ref{fig:systematic_errors}, we show the infidelity of a $\mathrm{CNOT}$ gate as a function of the systematic error in the shuttling speed and the waiting time.
We include the result for a static case, where the qubits are keep at a fixed distance $d=80$~nm, and there is a magnetic field gradient $\Delta B = 0.25$~mT/nm. 

\begin{figure}[t!]
	\centering
	\includegraphics{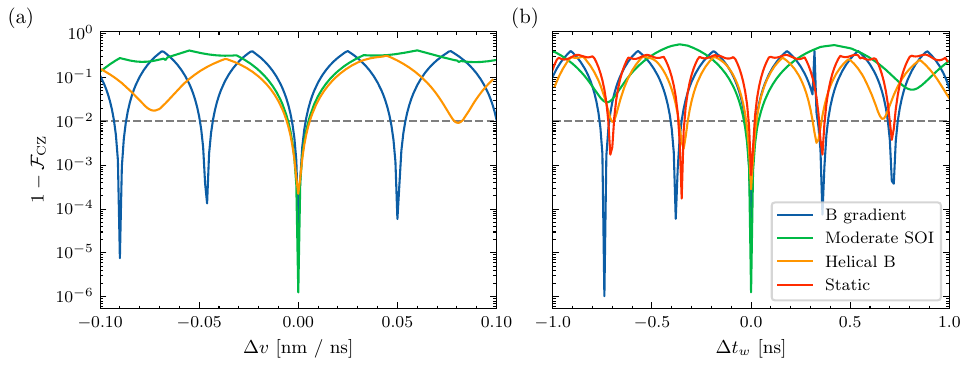}
	\caption{Infidelity of a $\mathrm{CNOT}$ gate as a function of (a) a systematic error in the shuttling speed, and (b) a systematic error in the waiting time.
	Colored lines correspond to different system configurations, as indicated in the legend.
	The horizontal dashed line indicates the threshold for $\mathcal{F} = 0.99$.
	By definition, there is no systematic error in the shuttling speed for the static case.
	}
	\label{fig:systematic_errors}
\end{figure}

The systematic error sensitivity is defined \cite{GueryOdelin2019} as the curvature of the fidelity as a function of the systematic error, i.e.,
\begin{equation}
	\mathcal{S}_v = \left.\frac{\partial^2 \mathcal{F}}{\partial \Delta v^2}\right|_{\Delta v=0}.
\end{equation}
The systematic error sensitivity to the waiting time is defined similarly.
Lower values of $\mathcal{S}$ indicate a more robust protocol against systematic errors.
In the following table we summarize the systematic error sensitivity for the different configurations studied in the main text.
\begin{table}[h!]
\centering
\setlength{\tabcolsep}{20pt} 
\renewcommand{\arraystretch}{1.1} 
\caption{Summary of the systematic error sensitivity for the shuttling speed $\mathcal{S}_v$ and the waiting time $\mathcal{S}_{t_w}$ for different protocols.
Lower values indicate a more robust protocol against systematic errors.}
\begin{tabular}{|c|c|c|}
\hline
\textbf{Protocol} & $\mathcal{S}_v$ [(ns / nm)$^2$] & $\mathcal{S}_{t_w}$ [(1 / ns)$^2$] \\
\hline
B gradient & 1667 & 28.7 \\
Moderate SOI & 763 & 15.6 \\
Helical B & 587 & 29.5 \\
Static & - & 90.2 \\
\hline
\end{tabular}
\label{tab:Sv_St}
\end{table}

We find that implementing a moderate SOI, or a helical magnetic field, we decrease the sensitivity to systematic errors in the shuttling speed, compared to the case of a linear magnetic field gradient.
Furthermore, the sensitivity in all these cases is lower than for a static qubit configurations.
These results indicate that the protocols studied in the main text are robust against systematic errors in the shuttling speed and waiting time, even improving the robustness of configurations which have already been experimentally achieved \cite{Matsumoto2025}.
Lower sensitivity to systematic errors can be achieved by decreasing $J_0(d_\mathrm{min})$, at the cost of a longer protocol time.

\section{Gate fidelity for incoherent dynamics} \label{sec:gate_fidelity_incoherent}
To model the effect of incoherent dynamics, we consider a Lindblad master equation of the form
\begin{equation}
	\frac{d\rho}{dt} = -i[H(t), \rho] + \sum_i \left(L_i \rho L_i^\dagger - \frac{1}{2}\left\{L_i^\dagger L_i, \rho\right\}\right),
	\label{eq:lindblad_master_equation}
\end{equation}
where $H(t)$ is the effective Hamiltonian, $\rho$ is the density matrix of the system, and $L_i$ are the Lindblad operators.
In our case, we consider a pure dephasing model, where the Lindblad operators are given by
\begin{equation}
	L_i = \sqrt{\frac{1}{2T_2}} \sigma_z^{(i)},
\end{equation}
with $T_2$ being the dephasing time, and $\sigma_z^{(i)}$ is the Pauli matrix acting on the $i$-th qubit.

A quantum channel is a map between a quantum state $\rho(t=0)$ and the output state $\mathcal{E}(\rho(t=0)) = \rho(t=\mathcal{T})$, due to the action of a Hamiltonian $H(t)$, together with any measurement or noise processes.
We reconstruct the quantum channel by numerically solving the Lindblad master equation.
Then, we compute the Pauli transfer matrices \cite{Wood2011,Hantzko2025} both for the target gate $U_T$, and for the quantum channel.
In general, the elements of the Pauli transfer matrix $\mathcal{P}$ of an $n$-qubit quantum channel $\mathcal{E}$ is given by, 
\begin{equation}
	\mathcal{P}_{i, j} = \frac{1}{2^n}\Tr\left[P_i \mathcal{E}(P_j)\right],
\end{equation}
where $P_i$ is the Pauli basis, with a total of $4^2=16$ elements in our two-qubit system.

To remove the action of local gates, we compute a modified target gate $\tilde{U}_T$, performing two general one-qubit gates on each qubit, such that
\begin{equation}
	\tilde{U}_T = L \cdot U_T \cdot R,
\end{equation}
with $L,R\in \mathrm{SU}(2)$.
The local gates can be written as $L=L_1\otimes L_2$ and $R=R_1\otimes R_2$, with each gate acting on the $i$-th qubit.
Each local gate is defined by three angles $\vartheta_i$, summing up to a total of 12 free parameters.
Then, we compute the fidelity between the quantum channel and the modified target gate as
\begin{equation}
	\mathcal{F}_{\vartheta_i}(\mathcal{E}, U_T) = \frac{\Tr\left(\mathcal{P}_\mathcal{E} \mathcal{P}_{\tilde{U}_T}^\dagger\right)}{\sqrt{\Tr\left(\mathcal{P}_\mathcal{E} \mathcal{P}_\mathcal{E}^\dagger\right)\Tr\left(\mathcal{P}_{\tilde{U}_T}\mathcal{P}_{\tilde{U}_T}^\dagger\right)}}.
	\label{eq:incoherent_fidelity}
\end{equation}
The above expression is a measure of the cosine similarity between two Pauli transfer matrices.
Finally, we optimize the angles $\vartheta_i$ to maximize the fidelity $\mathcal{F} = \max_{\vartheta_i} \mathcal{F}_{\vartheta_i}(\mathcal{E}, U_T)$.
To find the global maximum of the fidelity, we use a dual annealing algorithm \cite{Xiang1997}.
It is worth noting that due to the nature of the optimization algorithm, we are always obtaining a lower bound for the fidelity.

In Fig.~\ref{fig:pure_dephasing}, we show the fidelity of the $\sqrt{\mathrm{SWAP}}$ gate as a function of the dephasing time $T_2$, for different configurations.
We have chosen the shuttling speed and the waiting time such that the gate is achieved with high fidelity in the absence of dephasing, and the total protocol time is $\mathcal{T} \sim 80$~ns, much shorter than typical dephasing time of few $\mu$s.
The results obtained for the case of constant magnetic field and no SOI, and the results obtained for constant SOI are similar.
We have also included as a benchmark the case of static qubits in absence of SOI and constant magnetic field.
Here, the two particles are fixed at a given distance such that $J(d) \mathcal{T} / \hbar = \pi / 4 + k\pi$.
The results are again similar to the previous cases.
However, when including a helical magnetic field, the robustness to pure dephasing is significantly enhanced.
This enhancement is seen as a lower curvature of the fidelity as a function of the dephasing time $T_2$.
These results remark the benefits of implementing a helical magnetic field or strong SOI to both achieve more gates with high fidelity, and to enhance the robustness of the gates against pure dephasing.

\begin{figure}[t!]
	\centering
	\includegraphics{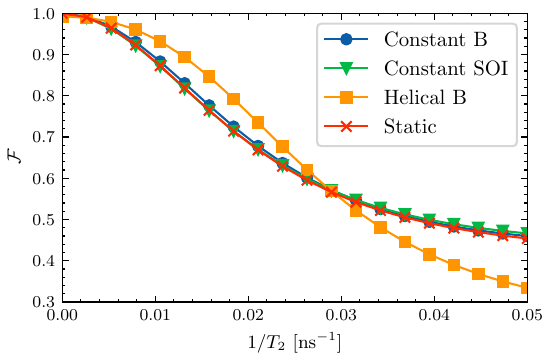}
	\caption{Fidelity of the $\sqrt{\mathrm{SWAP}}$ gate as a function of the dephasing time $T_2$, for different configurations, as shown in the legend.
	For all configuration, we have chosen the shuttling speed and the waiting time such that the total protocol time is $\mathcal{T} \sim 80$~ns.
	The static case is obtained from the case of constant magnetic field and no SOI, with $v=0$ and a distance between the two particles $d = 135.3$~nm.}
	\label{fig:pure_dephasing}
\end{figure}


\section{Filter functions}

To compute the effect of different noise sources on the gate fidelity, we use filter functions \cite{Green2013,Cerfontaine2021,Burkard2023,Hansen2023}.
Here, the error produced by a given noise source reads
\begin{equation}
	1 - \mathcal{F} \sim \int df S_i(f) F_i(f),
\end{equation}
where $S_i(f)$ is the power spectral density of the noise source, and $F_i(f)$ is the filter function.
The filter function is defined as
\begin{equation}
	F_i(f) = \sum_j\abs{\tilde{R}_{i,j}(f)}^2,
\end{equation}
\begin{equation}
	\tilde{R}_{i,j}(f) = \int_0^\mathcal{T}dt \Tr[U_c^\dagger(t) P_i U_c(t) P_j] e^{2\pi i f t},
\end{equation}
where $U_c(t)$ is the unitary time evolution operator in absence of noise.
For semiconductor quantum dots, the most relevant noise source is charge noise, which can be modeled as a low-frequency $S_i\sim 1/f$ noise.
In Fig.~\ref{fig:filer_functions}, we show the filter functions for the different Pauli string and for different system configurations.
Since our system is symmetric in the interchange of both quantum dots, the filter functions for the Pauli strings $P_iP_j$ and $P_jP_i$ are equal, so we only show one of them.
In all cases, the most relevant Pauli string is $\mathrm{IZ}$, which correspond to dephasing of a single qubit.
Here, we can see that the implementation of a static Hamiltonian result in a stronger sensitivity to pure dephasing.
The same is true for the case of the $\mathrm{XY}$ Pauli string, the second most relevant Pauli string.

\begin{figure}[t!]
	\centering
	\includegraphics{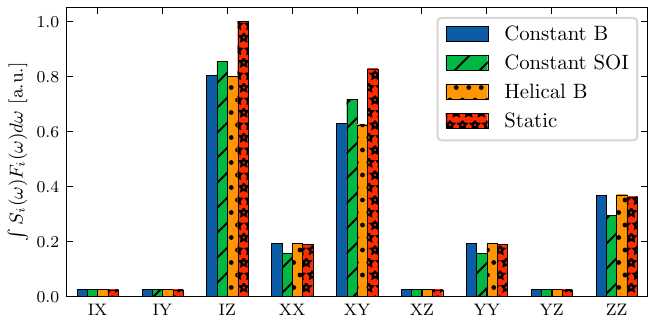}
	\caption{Filter functions for the different Pauli strings, with each system configuration shown in the legend.
	The target gate is $\sqrt{\mathrm{SWAP}}$.
	Other parameters are the same as in Fig.~\ref{fig:pure_dephasing}.}
	\label{fig:filer_functions}
\end{figure}

\end{widetext}

\bibliography{references}

\end{document}